# Force Matching and Iterative Boltzmann Inversion Coarse Grained Force Fields for ZIF-8

Cecilia M. S. Alvares,[1] Rocio Semino[2]*

[1] ICGM, Univ. Montpellier, CNRS, ENSCM, Montpellier, France
[2] Sorbonne Université, CNRS, Physico-chimie des Electrolytes et Nanosystèmes Interfaciaux, PHENIX, F-75005 Paris, France
* rocio.semino@sorbonne-universite.fr

ABSTRACT

Despite the intense activity at the electronic and atomistic resolutions, coarse grained (CG) modeling of MOFs remains largely unexplored. One of the main reasons for this is the lack of adequate CG force fields. In this work, we present Iterative Boltzmann Inversion (IBI) and Force Matching (FM) force fields for modeling ZIF-8 in three different coarse grained resolutions. Their ability of reproducing structure, elastic tensor and thermal expansion is evaluated and compared with that of MARTINI force-fields considered in previous work.[C. M. S. Alvares et al, J. Chem. Phys., 158, 194107 (2023).] Moreover, MARTINI and FM are evaluated in their ability of depicting the swing effect, a subtle phase transition ZIF-8 undergoes when loaded with guest molecules. Overall, we found that all our force fields reproduce structure reasonably well. Elastic constants and volume expansion results are analyzed and the technical and conceptual challenges in reproducing them are explained. Force matching exhibits promising results for capturing the swing effect. This is the first time these CG methods, widely applied in polymer and biomolecules communities, are deployed to model porous solids. We highlight the challenges of fitting CG force fields for these materials. This work opens the door to a whole new line of developments in the field of modeling MOFs and other porous crystalline solids.

I) INTRODUCTION

Simulations in the atomistic (AA) scale are essential for understanding structure, dynamics and physical properties at the microscopic level, as they provide information that cannot be experimentally resolved.[1-5] Yet, some properties or events cannot be adequately studied with these methods, since atomistic simulations become computationally prohibitive as the number of atoms increases significantly. Bridging the gap between macroscopic modeling approaches and atomistic simulations lies particle-based coarse graining (CG). In CG, groups of atoms are replaced by new entities, namely superatoms or beads, whose dynamics can be studied within Newtonian or Langevin mechanics by using suitable force fields to model the interactions. The use of CG models has become commonplace in the bio and polymer communities,[6-14] but the materials science community has not yet incorporated them to their routinely-used set of techniques. One of the key reasons for this is that the most popular CG potential fitting approaches, which include MARTINI, Force Matching (FM) and Iterative Boltzmann Inversion (IBI),[15,16,17,18] have been developed and mostly tested for molecule-based systems, and not for extended materials.

In this contribution, we develop and critically compare CG force fields for metal-organic frameworks (MOFs). MOFs are extended, generally crystalline, porous materials that result from the polymerization of metallic cations or oxoclusters with organic ligands. Their attractivity lies in the myriad possibilities for encapsulation, transport, selective separation and nano-reactor capacities offered by their connected porosity. Some of these materials contain mesopores, while others are microporous, and both can exhibit meso- or macro-porosities when in contact with other materials within mixed-matrix membranes. They can present large scale defects or show collective-long range motions that affect their porosity (i.e., the "breathing effect"). To model these phenomena, large simulation cells are required and CG models become a must. In this context, a systematic study of CG force field fitting strategies applied to MOFs is a much needed first step to drive future research directions. Are the most widely used strategies for fitting CG models appropriate for MOFs? What are their merits and setbacks? To answer these questions, we evaluate FM and IBI CG potential fitting strategies and compare them to previously derived MARTINI force fields to model ZIF-8[19] in coarsened resolution according to three different AA-to-CG mappings.[20] This work is the first systematic exploration of CG strategies applied to modeling MOFs, and it follows up two previous studies: one relying on MARTINI force fields and the other, on a genetic algorithm optimization of the Hessian of an atomistic benchmark.[20,21] ZIF-8 was selected because it is a widely studied material and it features excellent chemical and thermal stability as well as promising applications in health and environment domains.[22-24]

We compare FM, IBI and MARTINI models in their ability to reproduce structural (radial distribution functions, bond and angle distributions, lattice constants) and thermodynamic (elastic tensor, volume expansion) properties. Specifically for FM and MARTINI models, the ability of depicting a dynamic property (the swing effect, a collective motion-induced phase transformation experimentally observed in ZIF-8 when loading the pores with guest molecules)[25-28] is also investigated for the least coarsened mapping. We found that all our CG force fields reproduce well the structure of the materials. Thermodynamic properties are more complicated and they may depend on some technical aspects such as the quality and kind of pressure correction or the extrapolation scheme of the potentials in non-sampled regions of the phase space. Finally, FM qualitatively exhibits the swing effect.

This article is organized as follows. The theoretical foundations of the CG force fields necessary to the understanding of our results are summarized in Sec. II. Sec. III contains the methodological details, while Sec. IV and V feature the results and conclusion respectively.

II) THEORETICAL FOUNDATIONS

MARTINI force fields, developed primarily in the context of biomolecules and polymers, assume pre-defined analytical forms for the potentials and the parametrization for non-bonded interactions is determined aiming to reproduce thermodynamic properties in a general fashion so that many different systems can be described using a single set of bead flavors.[15,16] More specifically, in MARTINI 2, the parametrized force fields allow to reasonably reproduce the partitioning of different molecules in oil/water phases.[15] Recently, a more refined parametrization of MARTINI was developed incorporating the interaction energy between different molecules and a solvent and solvation free energy among other thermodynamic quantities as additional fitting targets.[16] Moreover, the ionic interactions were more carefully parametrized and a broader spectrum of CG resolutions and some general chemical specificities (e.g., capacity for $\pi$-$\pi$ interactions) were included. The parametrization of bonded potentials on the other hand allows depicting the spatial arrangement of the system.[15]

Contrary to MARTINI, force matching (FM) is a bottom-up method.[17,29,30] Essentially, it requires knowledge on the force experienced by the atoms in different configurations of an atomistic benchmark system. This implies that a reliable all-atom model is required *a priori*. Once the decision for a mapping is made, atoms can be lumped into their corresponding bead and the forces experienced by each of the beads, $F_I$, can be calculated from the force experienced by the atoms, $f_i$, that compose it. These reference forces experienced by the beads in the CG resolution serve as reference for fitting potentials that are defined for modeling the interactions. The fitting is achieved by finding the optimal solution for an overdetermined system of equations where the combined effect of the defined potentials will try to overall yield for each bead in each configuration a value of force as close as possible to the reference one it should be experiencing.

In principle, the notion of rightful reference forces that should be experienced in the CG resolution is not defined and thus different ways to calculate the reference forces $F_I$ from the atomistic forces are possible within the FM algorithm.[30] Very interestingly, works aiming to bridge the statistical mechanics in the AA and CG resolutions have proven that the sum of atomistic forces is a suitable choice for the reference force experienced by the beads.[31-33] If this definition of the force is applied to a mapping where beads are centered in the center of mass and no atoms are shared between beads, then the usual force-matching algorithm, in the ideal scenario of no residual error, leads to a Hamiltonian whose underlying probability density function in the CG level matches the corresponding one in the all-atom resolution. In such an ideal scenario, the CG model is said to be consistent with the underlying all-atom model. Notably, along the mathematical formalism for bridging the statistical mechanics of the all-atom and coarse grained resolutions, it has also been shown that, at each CG configuration, the Hamiltonian in the CG resolution that leads to consistency is related to the integration of the all-atom probability density function in a specific region of the all-atom phase space. Furthermore, if in each reference atomistic configuration the force on each CG bead is a linear combination of the forces experienced by the atoms forming it, the consistent Hamiltonian in the CG resolution is the equivalent of a constrained potential of mean force in a phase space defined combining atomistic and coarse grained degrees of freedom.

Lastly, iterative Boltzmann inversion (IBI) is a potential fitting strategy to derive CG force-fields that best reproduce the structure of the atomistic benchmark system.[18] Its algorithm picks up from the reversible work theorem, in which an analytical expression for the reversible work for bringing two particles at a given separation distance is determined.[34] The expression for the reversible work, that would rather be associated with the variation in Helmholtz free energy in the context of a transformation at constant (N,V,T), is used within the IBI algorithm as the initial guess for the potential modeling the interaction between the corresponding CG beads. Starting from initial guesses, IBI prescribes a formula to iteratively update the potentials such that an optimal structural description is achieved. The mathematical formula for the potential update in the iterative procedure should overall lead to the softening or sharpening of potentials if peaks in the distributions are sharper or softer, respectively, compared to the reference structure. Note that the iterative procedure entails losing the direct link between the potential and the reversible work, as the initial guess is continuously modified.

Both consistent-FM and IBI correspond to coarse grained potentials that are potentials of mean force (PMF). Yet, it is interesting to note that the two PMFs are different in nature. The PMF that appears when bridging the statistical mechanics of a system in its AA and CG resolutions is related to the fact that the Hamiltonian in the CG level is tied to a constrained region of the all-atom phase space.[31,32] On the other hand, the PMF involved in IBI's formalism is related to the work required to bring a given set of superatoms to a specific relative spacing, and there is no underlying link with the region of the phase space spanned by the consistent CG system.[34]

Finally, it is worth mentioning that reproducing pressure at CG resolutions is a challenge under discussion in the CG community.[35-38] This can be easily rationalized by looking at the expression for pressure derived within statistical mechanics, which depends on position and forces experienced by the structural units.[39] Different pressure correction methodologies have been developed for IBI and FM methods.[18,35,40,37,32] For IBI, the most common pressure correction is the linear one presented in the original IBI publication.[18] For FM, ensuring that the pressure is correctly reproduced is achieved by either introducing a constraint in the algorithm (Virial constraint) or by introducing a volume dependent term in the Hamiltonian.[37,32]

III) METHODOLOGY

In the current work, ZIF-8 is modeled in the CG resolution according to three different mappings: A, B and C, shown in figure 1. The center of the beads lies in the center of mass (COM) of the corresponding group of atoms. Bond, angle and pairwise non-bonded potentials are considered for modeling the interactions. No potential for dihedrals or improper were employed. A bond is considered to exist whenever superatom pairs encapsulate atoms that are known to be chemically bonded to one another. Angle potentials are assumed to exist for all sequences of three bonded superatoms. In IBI and FM models, non-bonded forces are not applied between bonded 1-2 and 1-3 neighbors and their cutoff was set to be 15 Å.

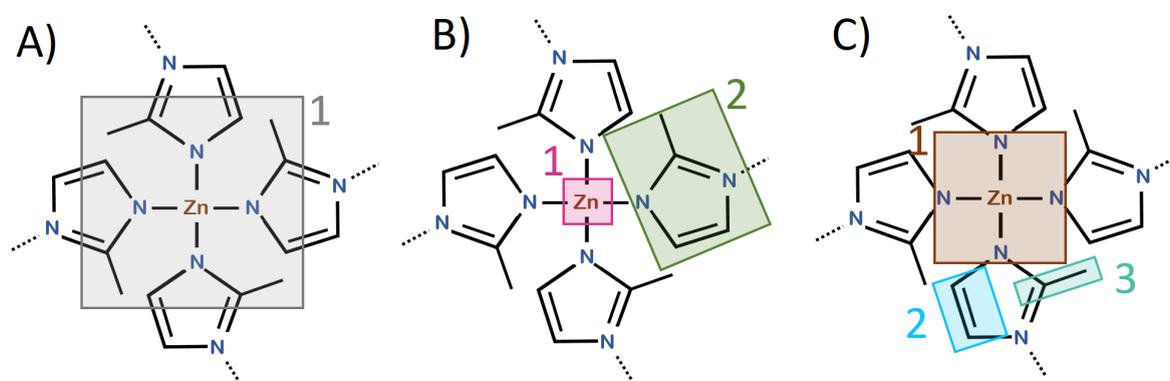

Figure 1. Mappings considered for studying ZIF-8 in coarsened resolution. (A) In mapping A, a single bead encapsulates a $Zn^{2+}$ ion together with half of the four bonded ligands. The -C-$CH_3$ group is shared between the two beads that encapsulate the same ligand. (B) In mapping B, each $Zn^{2+}$ ion is considered to be a whole bead, labeled type one, while each ligand is considered to be another bead, labeled as type two. (C) In mapping C, three bead types exist, with beads type one being the $Zn^{2+}$ ion together with the four closest nitrogens, beads type two being the -HCCH- group and bead type three being the -C-CH$_3$ group, as shown in the figure. These mappings were also investigated when developing MARTINI force fields in C. Alvares et al, J. Chem. Phys., 158, 194107 (2023).

For each mapping, two different models were created: one where the potentials were all derived via FM and the other, via IBI. MARTINI models for each of the mappings were taken from previous work.[20] Within the full spectrum of MARTINI models that we previously developed, we consider here models mAM2 II, mBM2 II and mCM2 II for comparison. All of them are based on MARTINI 2.0, since MARTINI 3.0 centers the beads on the center of geometry instead of the center of mass, and we aimed to be consistent in the bead-centering strategy for our comparison. Previous work has proved that the assigned bead flavor has very little influence over the results obtained, so the specific models were picked arbitrarily among the possibilities available considering polarity and chemical characteristics of the beads. The parameters of bonded potentials and the bead flavors for these specific models are presented in the Supplementary Material (SM). It is worth remembering that, contrary to FM and IBI, non-bonded interactions are considered for bonded 1-3 neighbors in all MARTINI models. Furthermore, the mapping C MARTINI model contains an additional bond potential regulating the distance between superatoms 2 and 3 belonging to the same ligand. The reader may refer to C. Alvares et al, J. Chem. Phys., 158, 194107 (2023) for more information.

All the all-atom and CG simulations were performed using LAMMPS.[41] In all cases, whenever dynamics is studied in the NVT and NPT conditions, Nose-Hoover equations of motions are used,[42] with damping constant values calculated as a function of timestep following LAMMPS' manual recommendation.[43] The exceptions are the simulations for pressure corrected FM models in the NPT condition, which are carried out using the MTK equations of motion as implemented in the BOCS package for handling the dynamics of volume dependent Hamiltonians.[44-46] The timesteps considered for the CG-MD simulations were 5 fs for mapping A, 2 fs for mapping B and 1 fs for mapping C for both FM and IBI models. Timestep was decreased for lower degrees of coarsening to avoid numerical instabilities during the dynamics.

A) Derivation of CG force fields

Classical MD simulations in the all-atom resolution are used to gather reference data to derive all the CG force fields. IBI requires reference structural information of the system in the due coarsened resolution, namely reference radial distribution functions (RDFs) and probability density functions for bonds and angles. These functions are used for the iterative procedure and for determining the initial guess of each potential. For FM, on the other hand, a representative amount of configurations each containing the corresponding force experienced by each atom, $f_i$, is needed.

To carry out the MD simulations in the all-atom resolution, the interactions in ZIF-8 were modeled by the ZIF-FF force field,[47] which consists of bonded and non-bonded contributions. Bonded contributions include harmonic bonds and angles and cosine-based dihedrals and impropers, while the non-bonded terms are given by Coulombic and 12-6 Lennard-Jones terms. The timestep considered for the all-atom simulations was 1 fs.

A.1 ) Iterative Boltzmann Inversion

To get structural data at T = 300K and P = 1 atm, MD simulations in the NPT ensemble were performed at the all-atom resolution and configurations of the equilibrated system were collected. A total of 1000 configurations, spaced by 500 fs, were considered for deriving structural data. Reference RDFs and probability density functions for angles and bonds were built using VOTCA.[30] The equilibrium lattice constant predicted by ZIF-FF at these conditions was of 16.96 Å, confirming previous work that affirms it to be slightly underestimated compared to the experimentally observed value (16.992 Å).[48]

Subsequently, initial guesses of the potentials were obtained using VOTCA to Boltzmann invert the respective RDFs and probability density functions. IBI's algorithm was carried out using in-house python codes interfaced with VOTCA at each iteration, which was used only for calculating the bonded probability density functions. The CG-MD simulations necessary at each iteration were performed using LAMMPS, being the RDFs an output of the simulation. The dynamics was carried out in the NVT ensemble, with the volume fixed to be the equilibrium one observed in the atomistic simulations (i.e., the volume corresponding to a lattice constant of 16.96 Å) and T = 300K. All bonded potentials were simultaneously updated in one iteration while the next iterations in the sequence were dedicated to updating only one non-bonded potential at each iteration. This scheme was cycled, giving rise to an alternating potential fitting approach and repeated until converging to force fields that satisfyingly reproduce the structure. We note that alternating the optimization of potentials within IBI is a common practice.[30,49] Further details about the code and methodology can be found at the SM.

All IBI models yielded pressures that were off, as previously observed for other systems.[18,35,50,51] To correct for it, the simple, *ad hoc* pressure correction proposed in the IBI publication was used,[18] which implies adding to the non-bonded potentials the term shown in equation (1). The value of the pre-factor *a* was adjusted for each mapping individually in order to attain the target pressure. For mappings B and C, which contain more than one non-bonded potential, the same pressure correction function was applied for all of them. The pressure correction did not have any significant impact in the ability of reproducing the structure at ambient conditions.

$$\Delta U(r) = a(1 - r/r_{cutoff}) \quad (1)$$

A.2) Force-matching

All FM potentials used to model the interactions are non analytical table potentials expressed as cubic splines. Bond potentials depend on pair distance, angle potentials on angle values formed by the three sequentially bonded beads and non-bonded interactions are modeled in pairwise fashion.

Deriving force fields via FM can be challenging for crystalline solids such as ZIF-8 depending on how the algorithm is implemented. Using non analytical table potentials requires a set of configurations in which superatoms interacting via each defined potential sample a continuous range of values of the variables they respectively depend on. This is required in order to duly find the optimal values of the forces the potential should yield across this continuous range. While some scientists advocate for constraining the value of the forces coming from a given potential to be zero at non-sampled regions,[29] not all softwares implement the FM algorithm allowing for such constraint, therefore continuous sampling becomes a must. This can be problematic for multiple peaked distributions in solid phases. Several strategies allow to counter this problem for condensed phase systems.[52,53] In the present work, simulations of ZIF-8 at ambient-conditions density but at 1500 K were made to obtain a continuous sampling. Ambient-conditions density is the one predicted by the atomistic model at T = 300 K and P = 1 atm. Note that ZIF-8 is not stable at T = 1500 K, but the harmonic bonded potentials in the all-atom model should guarantee that the crystal structure is kept. Verification of RDFs, ADFs and BDFs confirmed that the crystal structure was not damaged by the high temperatures, as only the desired broadening of the peaks was observed to occur.

The atomistic dynamics was considered in the NVT ensemble. Once the system equilibrated, a total of 4000 configurations containing information on forces experienced by the atoms were collected for the equilibrated system and used to derive force fields via FM. The configurations were sequentially spaced by 500 fs. Note that, within this approach, the FM model will be assumed to be transferable to (V = $V_{ambient}$, T = 300K) to study ZIF-8. Previous work has indicated that temperature transferability of FM force fields is easier to achieve than density transferability.[52]

Using the set of configurations previously mentioned, the FM forces associated to each potential were derived through VOTCA.[30] The reference CG force at each bead was calculated by summing the atomistic ones experienced by all atoms belonging to the given bead. For mapping A, this means that the shared atoms, namely the -C-CH$_3$ group, contribute to both beads they belong to. While this is not a problem within the FM algorithm *per se*, it does break the requirement for deriving CG models that lead to consistency within statistical mechanics.[31]

Once all the FM potentials are ready, an explicit volume dependent term for the Hamiltonian, $U_v(V)$, was derived for each of the three mappings aiming to correct the pressure. The explicit volume dependent term was assumed to have the analytical form proposed by Das and Andersen,[32] shown in equation (2), and truncated at n = 2 as often done in other works.[38,54] N is the number of superatoms and $\bar{v}$ is the average volume predicted by the atomistic model at the thermodynamic state for which $U_v(V)$ is being parametrized. Determination of the coefficients was done using BOCS.[55] While for mapping A the initial values of $\Psi_1$ and $\Psi_2$ yielded an $U_v(V)$ term that led to a close-enough value of equilibrium lattice constant at ambient (T,P) conditions, mappings B and C required applying an iterative method (often referred to as "self consistent pressure matching"). Yet, for mapping C, the iterative approach did not ultimately converge to satisfying results, thus, we considered the expansion truncated at n = 3 instead for this mapping. Finally, we note that the pressure correction was only used aiming to reproduce the average volume at ambient conditions, and not the instantaneous fluctuations that occur during the dynamics.

$$U_V(V) = \psi_1 N \frac{V}{\bar{v}} + \sum_{i=2}^{n} \psi_i N \left(\frac{V-\bar{v}}{\bar{v}}\right)^i \qquad (2)$$

B) Simulations in CG resolution

Once IBI and FM models were obtained for each mapping, they were evaluated in their capacity of reproducing structure, elastic constant and volume expansion. Results are compared to those coming from MARTINI models, which were taken from previous work.[20] Moreover, MARTINI and FM were also investigated in their capacity of reproducing the swing effect.

B1) Structure

The structure was evaluated in terms of RDFs, bonded probability density functions and lattice constant. The two former were determined for each model using data collected during MD simulations in the NVT ensemble at a thermodynamic state corresponding to (T = 300 K, V = $V_{ambient}$), where $V_{ambient}$ is the equilibrium volume predicted by the atomistic simulation at T = 300 K and P = 1 atm. During the simulation, 1000 microstates were saved and the structural data was obtained from it using VOTCA and LAMMPS.[30,41] The RDFs were built without 1-2 and 1-3 bonded neighbors. The equilibrium value of lattice constant was determined by doing MD simulations at the NPT ensemble in a thermodynamic state where (T = 300 K, P = 1 atm) and averaging the instantaneous values attained after the system has equilibrated. As IBI and FM models were pressure corrected, the thermodynamic states (T = 300 K, P = 1 atm) and (T = 300 K, V = $V_{ambient}$) should roughly be the same ones. Since MARTINI prescribes no pressure correction and no guarantee of reproducing the pressure exists *a priori*, these thermodynamic states are most likely not the same.

Simulations for computing structural data were also performed for the selected MARTINI models since the RDF data from previous work contained 1-2 and 1-3 neighbors and the lattice constant results in the previous work were derived at a lower pressure.

B2) Volume expansion

The volume expansion coefficient was determined by numerical approximation as shown in equation (3). The equilibrium volume at (300 K, 1 atm) and (272.5 K, 1 atm) were determined by averaging instantaneous values attained during the dynamics carried out in the NPT ensemble at those thermodynamic states, respectively.

$$\alpha_V [K^{-1}] = \frac{1}{V_{(300K, 1atm)}} \frac{V_{(300K, 1atm)} - V_{(300K, 1atm)}}{300 - 272.5} \qquad (3)$$

B3) Elastic tensor

Finally, for determining the elastic constants at (300 K, 0 GPa), twelve different strained states were considered. Six of them correspond to $\varepsilon_{xx}$ strains and the other six, to $\varepsilon_{xy}$ strains. Since ZIF-8 is cubic, only $C_{11}$, $C_{12}$ and $C_{44}$ are relevant as the other elastic constants are expected to be either equal to $C_{11}$, $C_{12}$ or $C_{44}$ or zero. In the case of both normal and shear strains, the strained states correspond to deformations of (-0.6%, -0.4%, -0.2%, 0.2%, 0.4%, 0.6%) taken from the equilibrium value of the lattice constant at (300 K, 0 GPa). Simulations in the NVT ensemble were performed at each deformed state and the relevant elements of the stress tensor were determined by averaging instantaneous values attained once the system has equilibrated. The instantaneous values of $P_{ij}$ are output by LAMMPS. A stress-strain plot can then be made and the corresponding values of elastic constants, estimated from doing a linear fit. For determining $C_{11}$ and $C_{44}$, ($e_{xx}$, $P_{xx}$) and ($e_{xy}$, $P_{xy}$) stress-strain relationships were respectively considered. For $C_{12}$, both ($e_{xx}$, $P_{xx}$) and ($e_{xx}$, $P_{yy}$) were taken into account and the final value of the elastic constant reported was calculated by averaging the two individual ones. Notably, the stress tensor that can be calculated and output by LAMMPS is valid exclusively for Hamiltonians that do not carry volume dependency. Thus, for the FM models, which have a volume dependent Hamiltonian, it was necessary to manually add a term to the values of $P_{ij}$ that were output. The form of the term that needs to be added can be found elsewhere.[39]

B4) Swing effect

The capability of depicting the swing effect was evaluated for FM and MARTINI models. The swing effect is a structural change that some MOFs, including ZIF-8, have been observed to undergo when loaded with gas above a given threshold of adsorbed molecules.[25,26] The new structure, also referred to as high-pressure (HP) structure, exhibits different bond lengths and equilibrium angles and is characterized primarily by a different degree of rotation of the methyl-imidazole linkers relative to the pore windows compared to what is observed in the empty framework at ambient pressure (AP).

The loaded state at which ZIF-8 adopts the HP structure varies depending on the guest as well as on temperature.[56] Grand Canonical Monte Carlo calculations of the ZIF-8 adsorption isotherm performed at 77 K for $N_2$ illustrated that the AP structure can explain the lower pressure range of the isotherm (approximately ≤20 $N_2$ molecules per unit cell) while the HP structure is needed to reproduce the higher pressure range (approximately ≥50 $N_2$ molecules per unit cell).[26] Aiming to see if FM and MARTINI models could predict the proper rotation of the imidazole linker in the CG resolution, MD simulations for ZIF-8 coarsened according to mapping C were performed for the empty and the loaded system with 51 $N_2$ molecules per unit cell. Contrary to previous work,[20] the $N_2$ molecules were modeled here in coarse grained resolution with a single bead representing the whole molecule. To model the interactions between the $N_2$ beads and between $N_2$ and the ZIF-8 superatoms within MARTINI, both C1 and SC1 flavors for the $N_2$ beads were investigated. Within FM, the potentials used for modeling the interactions between ZIF-8 superatoms are the ones discussed in section A.2 and two different strategies were set for deriving force matching potentials to model the $N_2$ interactions, both of them involving MD simulations of loaded ZIF-8 in the all-atom resolution. In the first strategy, namely FM-s1, the framework was initialized in the AP structure loaded with 19 $N_2$ molecules per unit cell, a loading at which the AP structure should remain as the stable one at 77 K. The system's volume was kept fixed so that the lattice constant of the crystal is 16.96 Å (which corresponds to that predicted by ZIF-FF for the empty framework at ambient conditions) and the dynamics was performed in the NVT ensemble. In the second strategy, namely FM-s2, ZIF-8 was kept fixed at the HP structure reported in previous work[25] and all interactions between ZIF-8 superatoms were turned off, while the gas molecules were allowed to move in the fixed NVT condition with the volume fixed at the one corresponding to the HP structure. A loading of 51 $N_2$ molecules per unit cell was considered in this case. The dynamics corresponded to a temperature of 300 K in both cases. This temperature was selected instead of 77 K to counter the lack of continuous sampling discussed in section A.2.

Within both strategies, the MOF and the $N_2$ interactions were modeled using ZIF-FF and TraPPe respectively,[47,57]] and the parameters of the LJ potential for the MOF-$N_2$ interactions were derived via Lorentz-Berthelot mixing rules. The simulation of both the empty and loaded framework in the atomistic resolution used as a benchmark for FM-s1 revealed the AP structure as the stable one. In fact, even when investigated at a loading of 51 $N_2$ molecules at 77K, in which the HP structure is experimentally known to be the stable one,[26] the AP structure is still found at the atomistic level.

A total of 1000 and 4000 configurations containing the forces experienced by atoms were saved for strategies FM-s1 and FM-s2, respectively, once the system has equilibrated. For FM-s1, all forces coming from interactions between ZIF-8 superatoms were discounted from the values of forces experienced by the ZIF-8 superatoms in all the 1000 configurations, leaving only the contributions coming from interactions with the guest. The configurations were then used as input for deriving the four non-bonded FM potentials for the $N_2$ interactions for each of the two strategies separately.

Subsequently, CG-MD simulations were carried out for the empty and loaded frameworks. For the simulations where ZIF-8 was loaded, both FM-s1 and FM-s2 were investigated and two distinct MD simulation setups were carried out for each of them, yielding a total of four simulations. In the first setup, ZIF-8 was initialized in the AP structure and the dynamics of the system in the CG resolution was carried out in the NVT ensemble with the volume fixed at the one corresponding to a lattice constant of 16.96 Å. In the second setup, the simulation box initially containing ZIF-8 in the AP structure was deformed at a steady rate of 0.0333 Å/fs in all directions up to the corresponding volume of the reported HP framework, corresponding to a lattice constant of 17.071 Å.[25] The goal of considering the latter approach was to see whether or not there was any improvement in depicting the swing effect when enabling such volume increase. The ability of depicting the rotation of the ligands was evaluated by comparing the histograms of the dihedral Zn-Zn-Zn-$CH_3$ (the so-called swing angle), defined in previous works.[58,20] The swing angle was considered in the CG resolution corresponding to mapping C as the angle formed by the beads encapsulating each of the four atoms forming the dihedral Zn-Zn-Zn-$CH_3$. Histograms for the swing angle were computed using an in-house python code and considering a total of 1000 configurations saved during the CG-MD simulations for the equilibrated system. It is worth mentioning that, in principle, the precise degree of rotation of the ligand and volume of the unit cell of the HP structure may vary depending on the guest and temperature of the system. Different HP structures have been found reported in the literature.[25,26] Here we consider the HP structure published by S. A. Moggach et al., Angew. Chem. Int. 48, 7087 (2009) as reference. Upon this consideration, the reference swing angle values for the AP and HP structure are ≈9° and ≈23°, respectively, for the CG resolution corresponding to mapping C.

## III) RESULTS AND DISCUSSION

### A) Structure

The structure of ZIF-8 was well reproduced for all CG models and mappings, with IBI outperforming MARTINI and FM as expected. Figure 2 shows a selected set of RDFs for mapping C in order to illustrate the performance of the three models for this mapping. The remaining structural data for mapping C as well as for the two other mappings can be found in the SM. The values of the lattice constants at (300 K, 1 atm) predicted by IBI, FM and two selected MARTINI models are shown in table 1 for the three mappings.

The lattice constant is well reproduced for both FM and IBI, as expected from the fact that they were pressure corrected. It is possible that the value predicted by the FM model for mapping B could be further improved by switching to an n=3 volume dependent term as was done for mapping C. In terms of reproducing the spatial arrangement, figures 2 and SM3-SM5 show that the IBI model almost perfectly matches the reference distributions within all mappings. The distributions obtained when modeling the system with FM models are quite good, as commonly observed in the literature.[17,59,38] This is particularly interesting considering that the force fields were derived for a high temperature thermodynamic state. Given that ideal FM models should have consistent probability density functions, they should also, as a natural consequence, be able to reproduce the structure. Therefore, any failure in doing so should come either from residual errors in the FM algorithm and/or from the fact that the FM force fields were derived at higher temperature.

MARTINI, which does not encapsulate the concept of pressure correction, also performs reasonably well. The distribution functions are also similar to the reference ones, which should be attributed to the fact that the harmonic bonded potentials are parametrized to keep the relative spacing between superatoms as in the crystal structure. As discussed in previous work,[20] the difficulty in maintaining a good structure description comes mainly when using MARTINI 2 models for very low coarsened ZIF-8 mappings, and it is tied to the overly repulsive non-bonded potentials which overcome the effect of the bonded ones. Importantly, at a thermodynamic state of (300 K, V = $V_{ambient}$), where $V_{ambient}$ is the volume corresponding to a lattice constant of 16.96 Å, MARTINI models predict a pressure that is quite off compared to the expected ≈1 atm value. For example, models mAM2II, mBM2II and mCM2II predict average pressures of -97 bar, -479 bar and 2050 bar, respectively, which means the dP/dV must have a correspondingly, sufficiently large magnitude to justify the fact that these models predict reasonably well the lattice constant at (300K, 1 atm).

It is interesting to note that IBI was successful in converging to force fields that accurately reproduce the overall structure even though ZIF-8 is a crystalline solid. The IBI method was developed in a context where the degrees of freedom, a word here used to denote the independent variables the potentials depend on, are roughly uncorrelated. This is afterall the reason why some systems allow for potentials to be iteratively optimized individually within IBI's algorithm without significant harm to the overall structural description. The significant correlation between degrees of freedom expected for ZIF-8 motivated the potentials to be optimized in alternating fashion. And even so, despite the set up, the alternating nature of the optimization does not ensure convergence to potentials capable of reproducing the structure, since in principle, updating one potential could continuously cause undesired changes in the overall spatial organization that would not be fixed as the algorithm goes on. In that sense, the success in converging to IBI models that reproduce the structure in the three mappings studied proves that it is possible to use IBI models in systems where degrees of freedom are not separable if the goal is to get a good structural description. Notably, it was also possible to carry out the standard pressure correction without compromising the structure.

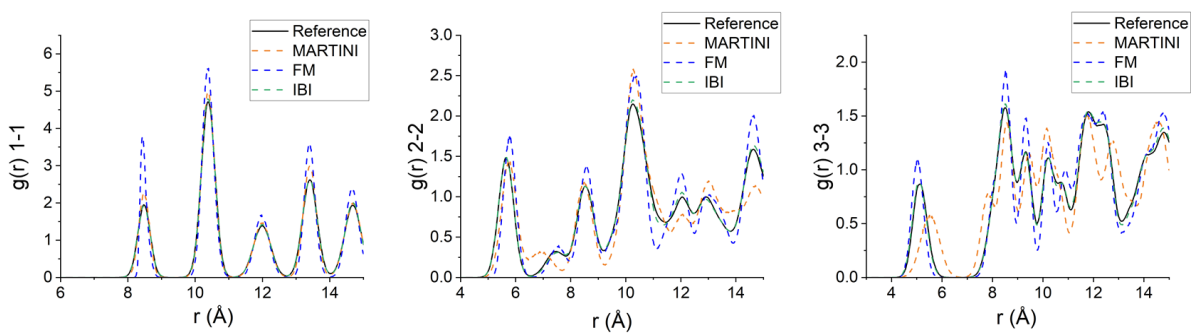

Figure 2. Selected RDFs: 1-1, 2-2 and 3-3 for mapping C modeled according to MARTINI, FM or IBI force fields. The 1-2 and 1-3 closest neighbors are not present in the plots. The MARTINI model is mCM2 II derived in previous work.[20] The reference curves were determined using coarsened all-atom configurations saved during the atomistic simulations.

B) Elastic constant

The elastic constants characterize the response of a system to a stress within the elastic regime.[60] The values of $C_{11}$, $C_{12}$ and $C_{44}$ for each model are reported in table 2 for IBI, FM and the selected MARTINI models of previous work.[20] It is interesting to note that in general MARTINI models outperform IBI and FM in reproducing elastic constants. This is surprising in the sense that MARTINI models were not parameterized to reproduce mechanical properties and were not even developed for MOFs.

When using IBI force fields to model interactions in a system, it is often found that the elastic constants are underestimated, leading to the common statement that IBI force fields are softer than they should be.[61-64] To improve the description of elastic constants, researchers have proposed either using specific equations of motion for the dynamics or adding special terms to the Hamiltonian for this purpose (although this can be detrimental to the ability to describe the structure).[65-67] Flags have also been raised to the fact that setting an arbitrary value for the cutoff of non-bonded interactions is not appropriate and that using larger $r_{cutoff}$ values should improve the ability of describing the pressure.[68] Yet, the empirical observation of the underestimation of $C_{ij}$ is not rooted in the IBI algorithm nor has any other theoretical foundation. In that sense, amongst the values of elastic constants obtained, the $C_{11}$ result obtained for mapping B can be considered a large outlier of the most commonly observed behavior.

The negative value of $C_{12}$ predicted for mapping A is more disconcerting. Elastic constants for ZIF-8 should be positive. Specifically in the case of applied tensile stresses and the corresponding strains, having positive values of elastic constants means that the force field responds favoring the shrinking in all directions. This response is expected, because as the tensile strain causes the distances between pairs to be larger than the equilibrium distances, shortening this distance in any direction would counter the perturbation. On the other hand, a negative value of $C_{12}$ suggests that the material favors expansion in the y and z directions in a $\varepsilon_{xx}$ deformed state. In the search for an explanation for the negative value observed, it is useful to remember that the algorithm of IBI is designed to work best in a scenario where degrees of freedom are essentially independent. In this scenario, IBI potentials exhibit a well whenever there is a peak in the corresponding distribution function. However, in the case of ZIF-8, bond lengths and angle values cannot be changed without significantly affecting the distribution of pairs interacting solely via the non-bonded potential.

The results show that the first iteration yields sharper distributions than prescribed by the reference for all mappings. In the ideal case of complete separability of degrees of freedom, this would mean that each of the potentials should be correspondingly smoothened as the IBI algorithm goes on. Yet, as the update in a potential affects the distribution associated to another due to the correlation of degrees of freedom, smoothening and sharpening can occur alternatingly in a non-predictable fashion as the algorithm progresses. For mapping A, convergence to a set of force field contributions that best describes the structure was achieved for a non-bonded 1-1 potential that counters the overly ordered arrangement imposed in the network by the bonded potentials, which force peaks in the 1-1 RDF to be sharper than they should. The non-bonded potential acquires as a consequence peaks within the range where there are peaks in the 1-1 RDF to alleviate the ordering, as opposed to the expected wells it should exhibit. The initial (step 1) potentials of mapping A and the corresponding target distributions for IBI's algorithm are shown in figure 3.

The profile of the 1-1 non-bonded potential leads the system to behave in the opposite direction to the one indicated by the reference distribution. If the superatom pairs are at larger distances than the equilibrium one, the potential yields forces that take them even farther apart. This unusual potential profile explains the exotic mechanical response to strains revealed by the negative $C_{12}$ value. For mappings B and C however, non-bonded potentials meet the expected profile IBI potentials should have in some ranges while not in others. The balance is such that positive elastic constants are obtained and the values predicted for mapping B and C are quite close to the reference ones. We can therefore conclude that while it is possible to converge to a pressure-corrected IBI force field that well reproduces the structure for a system with correlated degrees of freedom, caution should be taken if the goal is to reproduce the mechanical behavior. We thus advise checking the IBI potential profiles before blindly using them for this purpose.

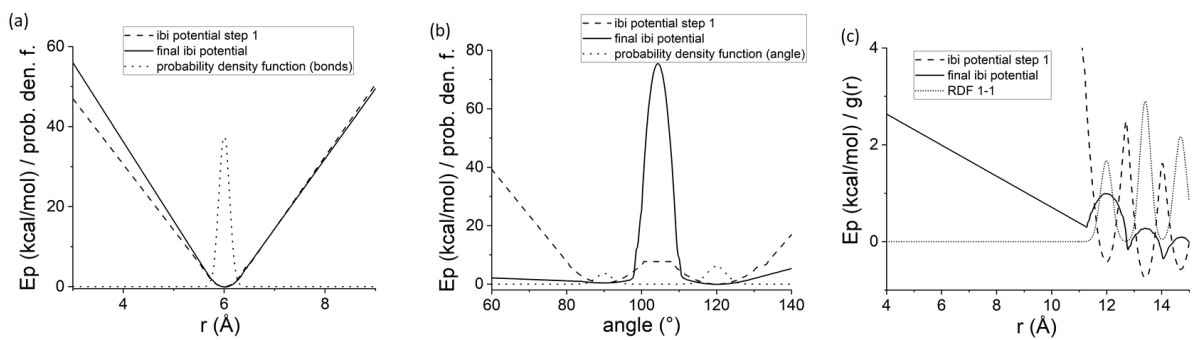

Figure 3. IBI potentials for step 1 and final step together with the corresponding reference bonded probability density (if bonded potential) or radial distribution function (if non-bonded potential) for mapping A.

Similarly, despite the reasonably good results obtained for mapping B and to some extent for mapping A, FM models also seem to have difficulties in reproducing the elastic behavior. In the present work, the values of the elastic constants obtained with by FM models pressure corrected with a $U_v(V)$ term were found to be very sensitive to the value of the $\Psi_i$ constants - much more than the equilibrium value of the lattice constant at ambient conditions. For example for mapping C, one of the $U_v(V)$ functions obtained at a given iteration of the self consistent pressure matching algorithm revealed a value of 14.40 GPa for $C_{11}$ and a non-linear relationship for $P_{xx} = f(\varepsilon_{xx})$, which had a $R^2$ of 0.39 for the linear fit. Furthermore, the negative $C_{44}$ value obtained for mapping C even breaks the condition for mechanical stability of cubic crystals.[69] It is possible that different choices for expressing the $U_v(V)$ term are more appropriate or that the Virial constraint may be more successful in ultimately converging to a Hamiltonian that reproduces the pressure in the context of MOFs. Indeed, there is no evidence that the pressure matching algorithm allows convergence to a Hamiltonian with actual physical meaning: it is rather an empirical correction to pressure and compressibility at the given thermodynamic state itself. Yet, the same can be said about the pressure correction via Virial constraint.[59] Ultimately, despite all the very interesting work that has been done so far in improving the ability of force-matching models to reproduce pressure-related properties, a physically meaningful approach, which may also enhance the ability of depicting mechanical properties, is still lacking.

C) Volume expansion

The values of volume expansion coefficient, $\alpha_V$, predicted by each model are also presented in table 1. Values of the selected MARTINI models calculated in previous works are also shown.[20] While the negative value predicted by the MARTINI model can be easily reasoned and predicted in the context of the analytical form of the potentials (see C. M. S. Alvares et al, J. Chem. Phys., 158, 194107 (2023)), this is not straightforward for FM and IBI force fields. Given the underlying physics of volume expansion within classical mechanics, changing the shape and/or steepness of the potentials in the region for which IBI and FM require extrapolation could lead to a model that better reproduces the value of $\alpha_V$. To prove this hypothesis, rough attempts to make the IBI potential for bonds sharper in its left hand-side for mapping A have shown to affect the volume expansion coefficient in the right direction. To achieve that, a sharper linear extrapolation was used and set to start at a larger value of $r$ for which IBI original data existed. Figure 4 shows a plot of the original IBI bond potential and the new one where repulsion was increased in the left-hand side. The non-bonded and angle potentials remained untouched. With such a rough approach, it was possible to increase the value of $\alpha_V$ from -156.6x10$^{-6}$ K$^{-1}$ to 9.5x10$^{-6}$ K$^{-1}$ at the cost of slightly damaging the structural description at (300 K, 1 atm) as revealed by a lattice constant increase of 0.8%. In principle, the same improvement could be achieved upon simply choosing a different form for the function used in the extrapolation. This shows that volume expansion is very sensitive to the onset of the potentials and, in turn, to the shape of the extrapolation function used. We thus suggest exploring changes in this direction to enable depicting the underlying physics of volume expansion phenomena by targeting the proper degrees of freedom.

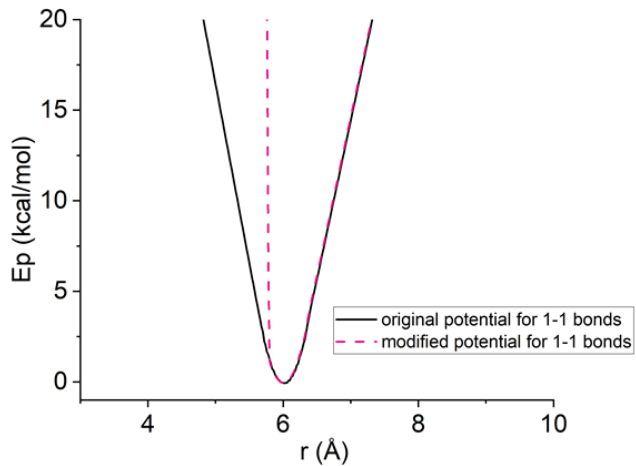

Figure 4 - Graphical representation of the original IBI potential for mapping A for 1-1 bonds and a modified version for the sake of improving the underlying value of thermal expansion coefficient predicted by the model.

|  | MARTINI | FM | IBI |
|---|---|---|---|
| mA | L = 16.95323 Å ± 0.00003 Å<br>$\alpha_V$ = -7.1E-6 K$^{-1}$ ± 0.3E-6 K$^{-1}$ | L = 16.94987 Å ± 0.00003 Å<br>$\alpha_V$ = -16.5E-6 ± 0.3E-6 | L = 16.96053 Å ± 0.00019 Å<br>$\alpha_V$ = -156.6E-6 K$^{-1}$ ± 0.2E-6 K$^{-1}$ |
| mB | L = 16.91593 Å ± 0.00009 Å<br>$\alpha_V$ = 40.7E-6 K$^{-1}$ ± 0.8E-6 K$^{-1}$ | L = 17.00407 ± 0.00005<br>$\alpha_V$ = -43.0E-6 ± 0.4E-6 | L = 16.96007 Å ± 0.00011 Å<br>$\alpha_V$ = 25.9E-6 K$^{-1}$ ± 0.1E-6 K$^{-1}$ |
| mC | L = 17.11197 Å ± 0.00007 Å<br>$\alpha_V$ = 21.7E-6 K$^{-1}$ ± 0.6E-6 K$^{-1}$ | L = 16.94493 ± 0.00007<br>$\alpha_V$ = 98.5E-6 K$^{-1}$ ± 0.6E-6 K$^{-1}$ | L = 16.95982 Å ± 0.00011 Å<br>$\alpha_V$ = 96.1E-6 K$^{-1}$ ± 0.1E-6 K$^{-1}$ |

Table 1. Values of lattice constant and volume expansion coefficient predicted by the MARTINI, FM and IBI models for mappings A, B and C.

|  | MARTINI | FM | IBI |
|---|---|---|---|
| mA | $C_{11}$ = 11.75 GPa<br>$C_{12}$ = 5.75 GPa<br>$C_{44}$ = 2.59 GPa | $C_{11}$ = 4.46 GPa<br>$C_{12}$ = 4.73 GPa<br>$C_{44}$ = 7.14 GPa | $C_{11}$ = 2.17 GPa<br>$C_{12}$ = -0.74 GPa<br>$C_{44}$ = 0.87 GPa |
| mB | $C_{11}$ = 10.07 GPa<br>$C_{12}$ = 4.3 GPa | $C_{11}$ = 13.55 GPa<br>$C_{12}$ = 3.44 GPa | $C_{11}$ = 17.49 GPa<br>$C_{12}$ = 7.94 GPa |
|   | $C_{44}$ = 1.41 GPa | $C_{44}$ = 2.48 GPa | $C_{44}$ = 1.72 GPa |
| mC | $C_{11}$ = 13.38 GPa<br>$C_{12}$ = 4.85 GPa<br>$C_{44}$ = 0.56 GPa | $C_{11}$ = 18.42<br>$C_{12}$ = 3.19<br>$C_{44}$ = -1.45 | $C_{11}$ = 6.19 GPa<br>$C_{12}$ = 2.46 GPa<br>$C_{44}$ = 1.61 GPa |

Table 2. Values of elastic constants predicted by the MARTINI, FM and IBI models for mappings A, B and C.

D) Swing effect

Histograms for the swing angle value for the empty and $N_2$-loaded ZIF-8 for MARTINI and FM models for mapping C are presented in figure 5. Negligible differences were observed in the CG-MD simulations of the loaded system at the AP structure volume versus the HP structure volume. Thus, only the structural results coming from the CG-MD simulation at the AP structure volume are presented in figure 5.

It is possible to see that already for the empty framework the MARTINI model significantly overestimates the due equilibrium value of the swing angle for the AP structure, whose reference value is of $\approx 9°$. Upon loading, the swing angle value increases for both $N_2$ bead flavors investigated within MARTINI. The mode lies at a higher value when considering $N_2$ to be of flavor C1 ($\approx 33°$ vs $\approx 29°$) and the peak is slightly broader, revealing the effect the non-bonded parametrization has in this context.

For FM, the model for the empty framework yields a swing angle value close to the one prescribed by the reference ($\approx 9°$). For the loaded framework, a shift of the peak of roughly 8° to the right is observed when using FM-s2 compared to FM-s1, with modes lying at $\approx 18°$ vs $\approx 10°$, respectively. This finding is very interesting taking into account the origins of the potentials used to model the $N_2$ interactions in the two cases. Using atomistic simulations where the framework is kept rigid at the HP structure and the gas is allowed to do its dynamics allows reaching FM potentials that lead to a ZIF-8 structure closer to the HP one in the CG resolution. This is very promising since the dynamics of the all-atom simulation used to derive the $N_2$ FM potentials has been carried out at a temperature higher than 77 K and the FM potentials modeling interactions between ZIF-8 superatoms were derived considering the AP structure only. It is also worth remembering that, ZIF-FF together with TraPPe were unable to predict the HP structure as the stable one when considering the framework to be flexible at the atomistic level, as mentioned in section B4. Given the improved structural results obtained when using FM-s2 compared to FM-s1 shown in figure 5, it is fair to say that despite being unable to reproduce the swing effect when considering the framework flexible, modeling the interactions with ZIF-FF together with TraPPe at least allows for the guest to have a dynamics and spatial arrangement that is closer to that of the HP structure. This seems to be already sufficient to lead to FM potentials for the $N_2$ interactions that significantly improve the capability of the model of depicting the swing effect in the CG level, showing the incredible potential of FM for capturing simple phase transitions.

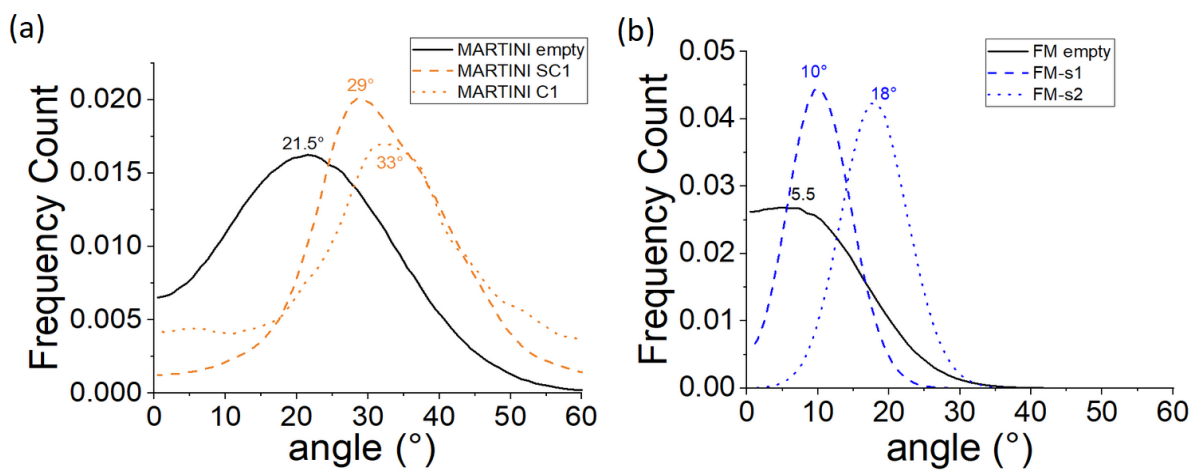

Figure 5 - Swing angle histograms for the equilibrated ZIF-8 structure loaded (dashed and dot lines) and empty (solid line) when interactions in the system are modeled with (a) MARTINI and (b) FM models.

CONCLUSION

In this work, MARTINI, FM and IBI models were studied in their ability to reproduce the structure and thermodynamic properties (elastic tensor and volume expansion coefficient) of ZIF-8. The capability of depicting the swing effect, a structural transition ZIF-8 undergoes upon guest adsorption, was also evaluated for MARTINI and FM models. We found that all CG strategies explored reproduce structural properties with a very reasonable accuracy. IBI models performed best in this task despite the fact that they are not developed to function in a context of high correlation between degrees of freedom, as is the case in a crystalline material.

In regards to thermodynamic properties, the results are varied. Elastic tensor is quite well reproduced by MARTINI 2.0 models. IBI was found to sometimes overestimate elastic constants, contrary to the underestimation tendency observed in the literature. The coarsest mapping exhibits a negative value for $C_{12}$, which has been tied to the fact that degrees of freedom in ZIF-8 are highly correlated. Despite the susceptibility of leading to models with anomalous mechanical behavior in some cases, IBI can also allow reaching models that properly depict it, as revealed for mappings B and C, for which the elastic constants were reasonably reproduced. FM models yield reasonable results for mappings A and B, but also presented challenges for reproducing elastic constants. Amongst the biggest deviations lies the negative $C_{44}$ predicted for mapping C. The results obtained are very sensitive to the $U_v(V)$ term used in the context of self-consistent pressure matching. Exploring the Virial constraint, using a different expression for the volume dependent term or developing more suitable ways of pressure correcting FM potentials are possible pathways for improving the description of the mechanical behavior of MOFs. Ultimately, we conclude that IBI and FM potentials should be used with caution when applied to depicting mechanical behavior of MOFs.

Concerning volume expansion, all strategies give overall reasonable results for the intermediate and less coarse mappings, while mapping A seems to be pathological in all cases. It is shown how the volume expansion prediction can be improved within IBI models by tailoring the onset of potentials accordingly. This is not recommended as a good practice by itself, but rather as an inspiration for ideas to enable the description of the proper physics underlying the volume expansion phenomenon.

Finally, we explored the possibility of reproducing the swing effect. While MARTINI force fields were not able to do so, we found that FM reproduces the effect to a large extent as the average values of swing angle for the empty and loaded framework are closer to the reference values. This is quite remarkable taking into account that reproducing the swing effect is already quite rare when relying on atomistic force fields and it is not achieved by the underlying atomistic benchmark model. The fact that the FM potentials between superatoms of the MOF were derived for the AP structure and for a higher temperature than the one for which the loaded state should yield an HP structure just adds up to the promising performance of FM in capturing simple phase transitions.

To the best of our knowledge, this article constitutes the first systematic study of coarse graining potential fitting strategies applied to MOFs, and it is also the first study to address the challenges of fitting CG potentials for porous solid materials. In this sense, these findings are not only valuable for the MOF simulation community, but also for researchers that focus on CG methods development. We hope our study motivates the wider use of CG models for MOFs and other materials.

ACKNOWLEDGEMENTS

The authors thank the École Doctoral Sciences Chimiques Balard for funding this work. R. S. thanks ERC Research for an ERG StG (MAGNIFY project, number 101042514). The authors are grateful to William Noid and Maria Lesniewski for helping with the usage of the BOCS code. This work was granted access to the HPC resources of CINES under the allocation A0130911989 made by GENCI.

DATA AVAILABILITY STATEMENT

A supplementary material supporting results presented in the paper is available for download. The FM and IBI force fields developed in this work can be found in https://github.com/rosemino/CG_ZIF-8 together with an example of the python codes used to perform IBI.

Supplementary Material for: Force Matching and Iterative Boltzmann Inversion Coarse Grained Force Fields for ZIF-8

Cecilia M. S. Alvares,[1] Rocio Semino[2]*

[1] ICGM, Univ. Montpellier, CNRS, ENSCM, Montpellier, France
[2] Sorbonne Université, CNRS, Physico-chimie des Electrolytes et Nanosystèmes Interfaciaux, PHENIX, F-75005 Paris, France
* rocio.semino@sorbonne-universite.fr

**Table of Contents:**

1. MARTINI

The MARTINI models used for comparing with Force-matching (FM) and Iterative Boltzmann Inversion (IBI) models were taken from previous work.[1] For mapping A, model mAM2II was considered, in which bead type one is of flavor $N_a$. For mapping B, model mBM2II, which assumes bead flavors P5 for both bead types 1 and 2, was selected. Finally, for mapping C, model mCM2II, which considers bead flavors P5, SC1 and SC1 for beads type 1, 2 and 3 respectively, was chosen. The bonded parameters for these models are presented below in table SM1. As a reminder, angles whose reference histogram accuses multiple peaks had one potential per peak. The non-bonded parameters can be found in the paper in which these models were developed for ZIF-8 or in MARTINI 2.0 official's publication.[1,2]

|  | Potential | Parameter values |
|---|---|---|
| Mapping A | bond 1-1 | $K_b$ = 29.05 kcal/mol ; $R_o$ = 6.0 Å |
|  | angle 111 | $K_a$ = 31.10 kcal/mol ; $\theta_o$ = 90.0°<br>$K_a$ = 42.77 kcal/mol ; $\theta_o$ = 120.5° |
| Mapping B | bond 1-2 | $K_b$ = 61.0 kcal/mol ; $R_o$ = 3.05 Å |
|  | angle 121 | $K_a$ = 540.0 kcal/mol ; $\theta_o$ = 160.4° |
|  | angle 212 | $K_a$ = 30.0 kcal/mol ; $\theta_o$ = 97.4°<br>$K_a$ = 48.0 kcal/mol ; $\theta_o$ = 115.7° |
| Mapping C | bond 1-2 | $K_b$ = 45.0 kcal/mol ; $R_o$ = 3.61 Å |
|  | bond 1-3 | $K_b$ = 18.0 kcal/mol ; $R_o$ = 3.18 Å |
|  | bond 2-3 | $K_b$ = 109.0 kcal/mol ; $R_o$ = 3.02 Å |
|  | angle 121 | $K_a$ = 23.5 kcal/mol ; $\theta_o$ = 112.3° |
|  | angle 131 | $K_a$ = 21.0 kcal/mol ; $\theta_o$ = 142.1° |
|  | angle 212 | $K_a$ = 12.0 kcal/mol ; $\theta_o$ = 102.0°<br>$K_a$ = 10.0 kcal/mol ; $\theta_o$ = 127.2° |
|  | angle 313 | $K_a$ = 0 kcal/mol ; $\theta_o$ = 80.22°<br>$K_a$ = 0 kcal/mol ; $\theta_o$ = 126.0° |
|  | angle 312 | $K_a$ = 240.0 kcal/mol ; $\theta_o$ = 52.7°<br>$K_a$ = 0 kcal/mol ; $\theta_o$ = 82.5°<br>$K_a$ = 0 kcal/mol ; $\theta_o$ = 147.8° |

Table SM1. Potential parameters of the MARTINI models reported in this work for comparison with FM and IBI models.

2. Iterative Boltzmann Inversion

The IBI algorithm was executed using in-house python codes interfaced with VOTCA,[3] which was used to generate probability density functions for bonds and angles at the iterations in which bonded potentials were modified. On the other hand, the RDFs used in the update of non-bonded potentials, were built during the dynamics using LAMMPS.[4] An example of code used for the IBI in a non-bonded potential can be found in https://github.com/rosemino/CG_ZIF-8.

Inspired by the work of M. P. Bernhardt et al.,[5] the code addresses potential problems that may either prevent the algorithm from converging to numerically stable simulations or significantly increase the noise on the potentials. It is not uncommon to have significantly small values in the onset regions of the peaks in distributions. Besides being very small, the magnitude of these values often oscillates in a random fashion. When coupled with the logarithm operation iteratively performed within the IBI algorithm, this feature can lead to unphysically high and steep values of potential energy, as well as to noisy curves. To prevent this issue from happening, the code scans the onset region of the distributions and zeroes any value below 0.005 before the first occurring one that reaches or surpasses this threshold. Furthermore, another possible source of noise may come from a non monotonic behavior in the onset region of the peaks of the distribution, which is a feature that can arise naturally as distributions are built from finite dynamics. To fix that, the code collects (x,y) points of the onset region of the original distribution with y values $\in$ (0.005, 0.1] and uses them as a basis for a cubic spline interpolation scheme for replacing the value of y for points having $y < 0.005$. In case the cubic spline interpolation setup fails to replace the $y$ value to ensure monotonic behavior, a different strategy is employed (see python code in the github). From a sample of empirical observations, points with $y > 0.1$ exhibit a monotonic behavior up to the mode of the corresponding peak and therefore these don't need to be modified. These thresholds are inspired in previous works[5] and, since the overall code allowed the IBI algorithm to successfully converge to potentials capable of reproducing the structure, they were kept.

Once the given distribution is pre-treated, the IBI algorithm is carried out for the corresponding potential. The update, ΔU, to be added to the potential shown in equation (SM1) specifically for the case of a non-bonded potential. Since values of $ln(0)$ are not defined, interpolations are performed to get values of the functions $k_B$Tln(g(r)$_{target}$) and $k_B$Tln(g(r)$_{model}$) in non-sampled intervals lying between peaks of the model and reference distributions. Second order polynomials are used for the interpolation and a smoothening was made where data from the interpolation meets data from the original functions. No extrapolation was done to gather data for the Boltzmann inverted distributions in non-sampled intervals that do not lie between peaks. Rather, the update ΔU was calculated only for intervals in which data either coming from the Boltzmann inversion or from the second order polynomial interpolation exist for both $k_B$Tln(g(r)$_{target}$) and $k_B$Tln(g(r)$_{model}$) simultaneously. The same holds for the bonded distributions. An illustrative sketch of the set up is presented in figure SM1 using as example the reference 312 angle probability distribution function for mapping C. A pre-factor of 0.25 or smaller was applied to ΔU before updating the current potential.

$$\Delta U(r) = k_B T \ln\big(g(r)_{target}\big) - k_B T \ln\big(g(r)_{model}\big) \qquad \text{(SM1)}$$

Finally, after the update, a set of points (x,y) on the onset of the updated potential for which the value of ΔU was defined, was used as a basis for a linear fit used for deriving new values in the non-updated range. These would be the ranges qualitatively indicated in green in figure SM1 as "No data for the Boltzmann inverted distribution". In the case of bonded potentials, such ranges should be the left hand-side until the first peak in the corresponding distribution and from the onset of the last peak until the last point for which data exists. In the case of non-bonded potentials, only the former range needs the extrapolation. Care is taken to always generate a linear extrapolation that is coherent with the expected physical behavior for a classical potential in terms of repulsion and attraction. Finally, a smoothening was made where data from the extrapolation meets data from the updated potential.

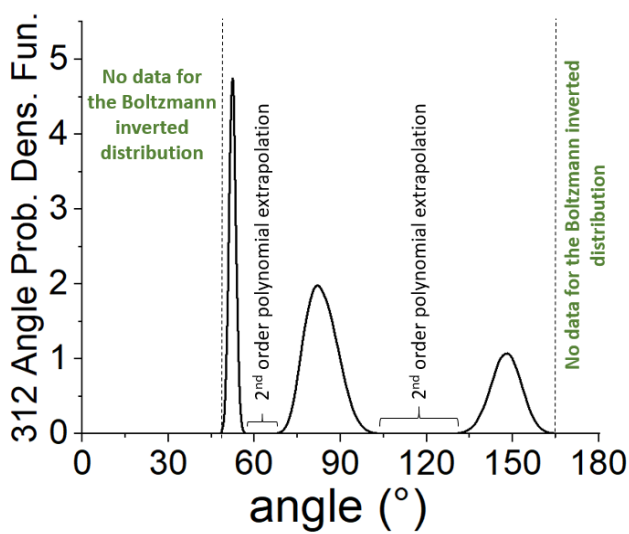

Figure SM1. Sketch of the set up used for computing the reference and model Boltzmann inverted distributions functions, required for computing the potential update at each iteration. The sketch was made using the 312 probability density function within mapping C as an illustrative example. A second order polynomial extrapolation is made for each of the Boltzmann inverted distributions (reference and model) individually before computing the ΔU function. A linear extrapolation is performed after the ΔU is added to the function, using values of the onset region of the updated potential in regions where values of ΔU were not defined.

The resulting IBI potentials are table potentials which were used within LAMMPS refining the bin size so that 10000 points exist for each of them (see pair_style table for more information).[6]

3. Force-Matching

*Potentials for interactions between superatoms of ZIF-8*

The force-matching algorithm was used to derive potentials at a thermodynamic state at ambient conditions density but at a temperature of 1500K. Upon comparing the structure at 300 K versus 1500 K in the different mappings investigated, it was possible to see that all peaks broaden to the point where continuous sampled regions were formed in all distributions except for angles 111 in mapping A. To counter the lack of sampling problem, a different potential was considered for angles belonging to each of the two peaks. For all mappings, bond and non-bonded potentials were set to have a bin size of 0.005 nm and angle potentials, of 0.001 radians.

It has been reported that dividing the overall overdetermined system of equations in smaller systems, solving them individually and averaging the solution yields force fields that more accurately describe the given system.[7] With that in mind, systems of equations built using 50 configurations each were considered for mappings A and B. For mapping C, 100 configurations per system of equations were considered.

The force-matching data output by VOTCA for each potential was smoothened and extrapolated in order to derive data that lies outside of the range of the continuously sampled one. Linear functions were used for the extrapolations and they were made so that all bond, angle and non-bonded potentials have values throughout the intervals of [0 nm, 1.2 nm], [0°, 180°] and [0.1 nm, 1.5 nm] respectively. Only points near the onset region for which original FM data exists were used as basis for fitting the linear function used for the extrapolation. After extrapolation, a smoothening was made in the interval where data coming from the extrapolation meets original data coming from the FM algorithm. In the case of non-bonded potentials, the range of sampling went up to the desired upper limit value (1.5 nm), also used as $r_{cutoff}$ for the non-bonded interactions in all cases, so only an extrapolation in the left-hand side of the curve was required. The non-bonded potentials were all shifted to have a value of 0 at $r = r_{cutoff}$.

It is interesting to note that the FM relies on a numerical optimization that does not necessarily yield physical potentials, which means that force values in the onset region may point to unphysical behavior if used as it originally is for the linear extrapolation. More specifically, the forces may lead to indefinite attraction as $r \to 0$ or $\theta \to 0$ or to indefinite repulsion as $r \to +\infty$ or $\theta \to 180°$. Care is always taken to not endorse such behavior whenever the original data causes it to occur. To achieve this, the sign of the slope of the optimal linear function obtained fitting the original data is inverted so that the potential accuses the proper expected physical behavior. An example is shown in figure SM2 for the 1-1 non-bonded potential of mapping B. The figure shows the region in which data obtained via extrapolation meets the data obtained with the FM algorithm, in which it is possible to see that the slope of the optimal linear fit was inverted in order to generate the linear extrapolation.

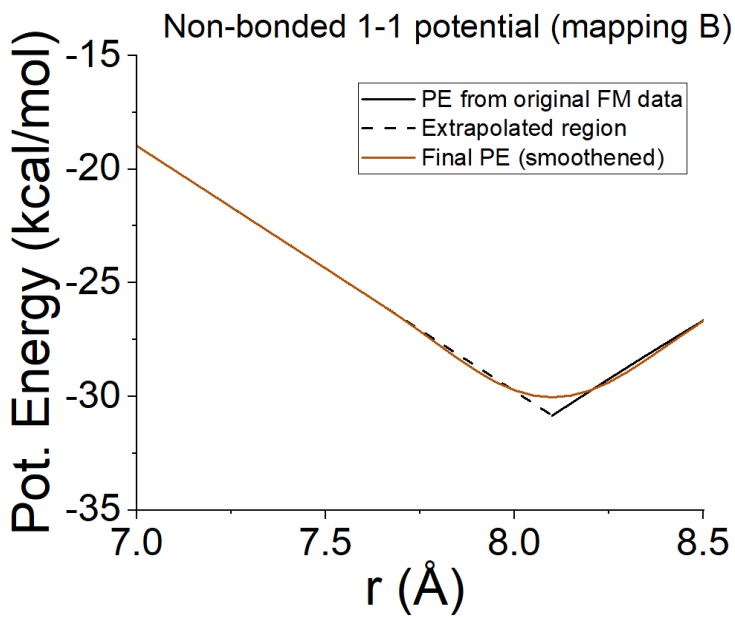

Figure SM2. Interval of potential energy (PE) values for the 1-1 non-bonded potential in mapping B in which data coming from FM algorithm (black solid line) meets data of the linear extrapolation used for deriving values in the non-sampled region (black dashed line). This is an example of a case where the optimal linear fit obtained with the PE data coming from FM algorithm had its slope sign inverted in order to ensure proper physical behavior at smaller values of $r$. The final potential curve, which was smoothened in the region where the data coming from FM algorithm meets the data coming from the extrapolation, is also shown in the plot (brown solid line).

Similarly to IBI, the resulting FM potentials are also table potentials, used within LAMMPS refining the bin size so that 10000 points exist for each of them (see pair_style table for more information).[6]

Further on, for the pressure correction, the target instantaneous volumes and corresponding pressures were taken from an MD simulation in the NPT ensemble at ambient conditions (T = 300 K, P = 1 atm) in the all-atom resolution, when parametrizing the volume dependent term.

*Potentials for N2-N2 and N2-(ZIF-8 superatoms) interactions*

In both strategies, the non-bonded potentials were built considering a bin size of 0.005 nm and the cutoff used for the non-bonded interactions were the same as the ones used for the interaction between ZIF-8 superatoms (15 Å). Each set of 100 configurations was used to build a different system of equations within the FM algorithm for each of the two strategies.

4. VALIDATION - Structural results

The RDFs and bond and angle probability density functions (BPDFs and APDFs) predicted by all models for all three mappings can be found in figures SM3 to SM5 together with the respective reference. The 1-2 and 1-3 bonded pairs are not shown in the RDFs. It is possible to see that the IBI models excel compared to FM and MARTINI models for all mappings investigated. FM models overall predict the due equilibrium positioning correctly but tend to overly restrain the movement, as indicated by the sharper and higher peaks. The success in reproducing structure supports temperature-transferability of the FM potentials, which had been derived at 1500 K. For mappings A and B, MARTINI models capture well the spatial arrangement of bonded superatoms, which is tied to the parametrization strategy for the bonded potentials. Small differences in peak height and width should come from the fact that these potentials have been parameterized in the previous work using angle and bond distribution functions built with a less refined bin size than in the current work. The difficulty in maintaining the description of bonded distributions within mapping C is linked to the non-bonded potentials prescribed within MARTINI 2.0 being overly repulsive, as discussed in previous work.[1] Despite not being able to keep up with the perfect description obtained using IBI models, ultimately the FM and MARTINI models were able to provide a reasonably good structural description for ZIF-8.

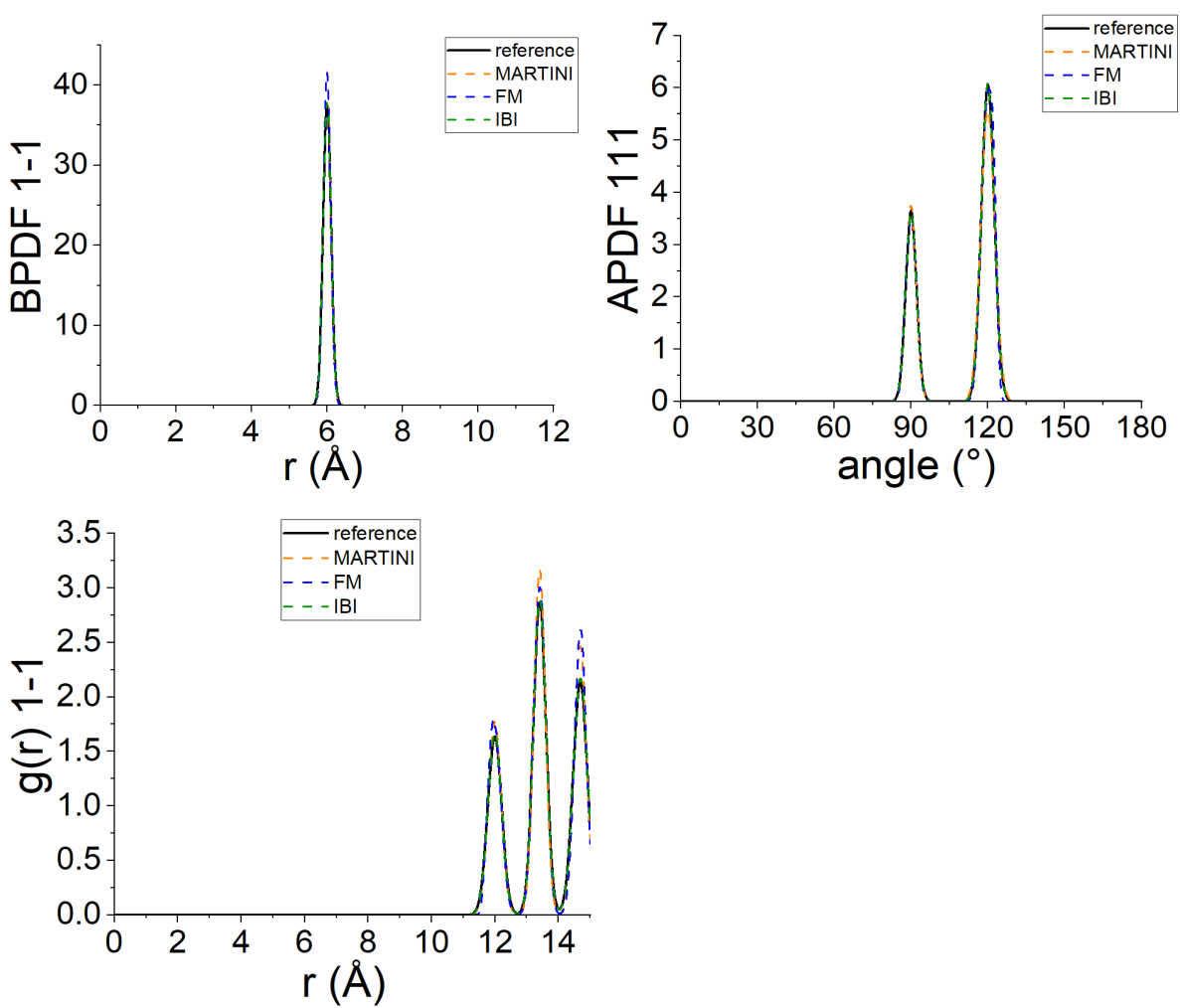

Figure SM3. Structural results given by MARTINI (dashed orange line), FM (dashed blue line) and IBI (dashed green line) models for mapping A. Bond and angle probability density functions are referred to as BPDF and APDF respectively. Reference distributions (solid black line) are also shown.

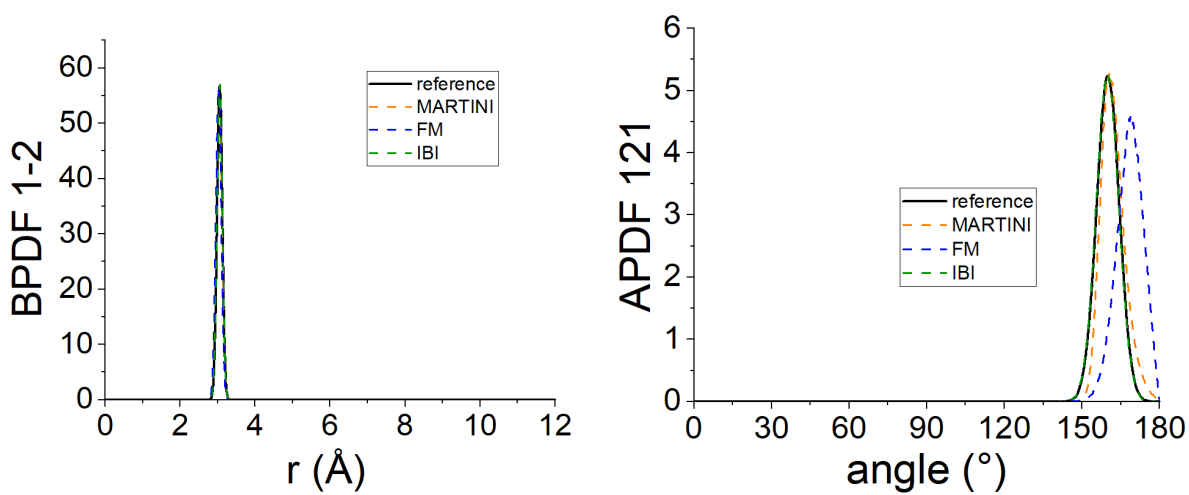

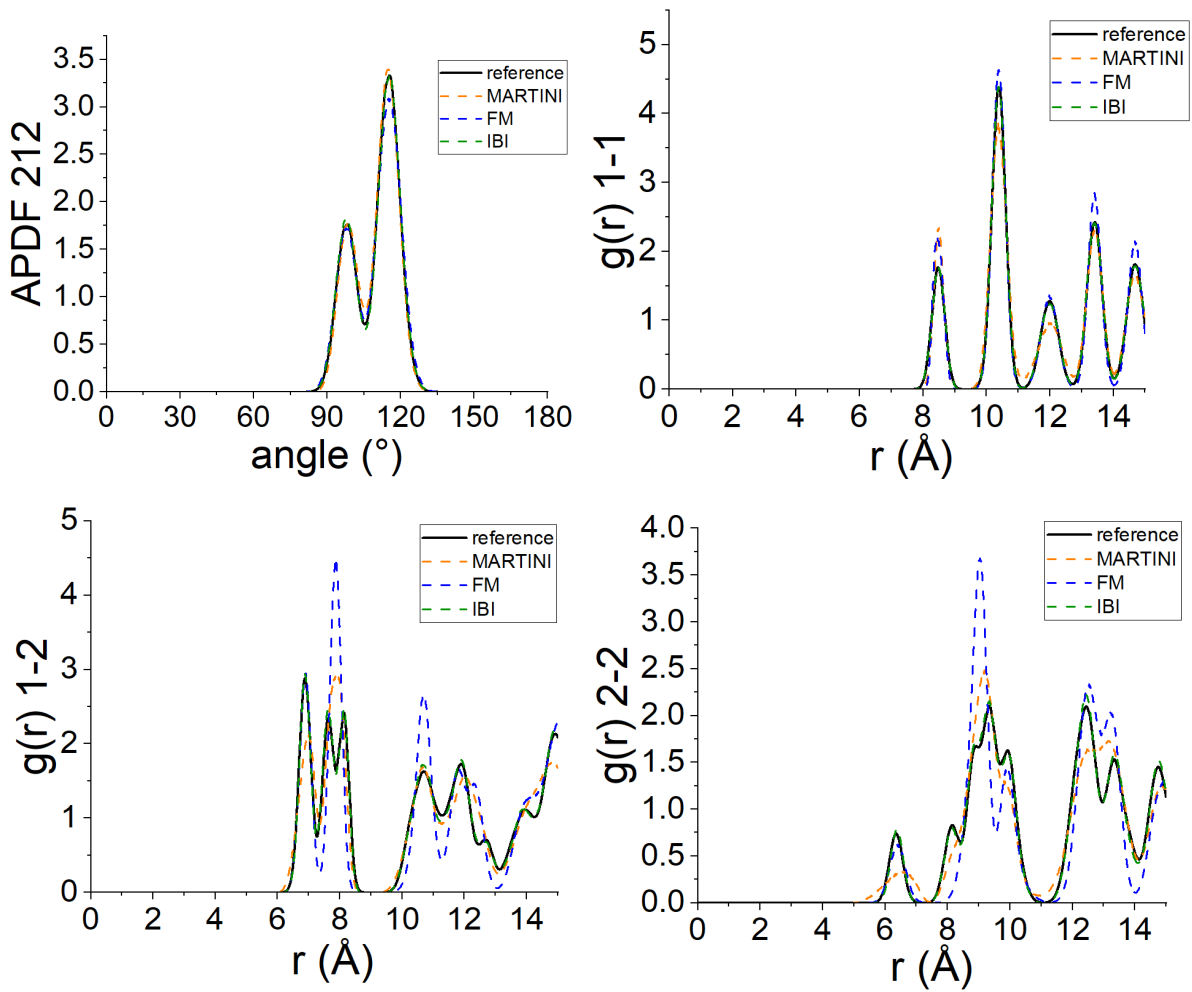

Figure SM4. Structural results given by MARTINI (dashed orange line), FM (dashed blue line) and IBI (dashed green line) models for mapping B. Bond and angle probability density functions are referred to as BPDF and APDF respectively. Reference distributions (solid black line) are also shown.

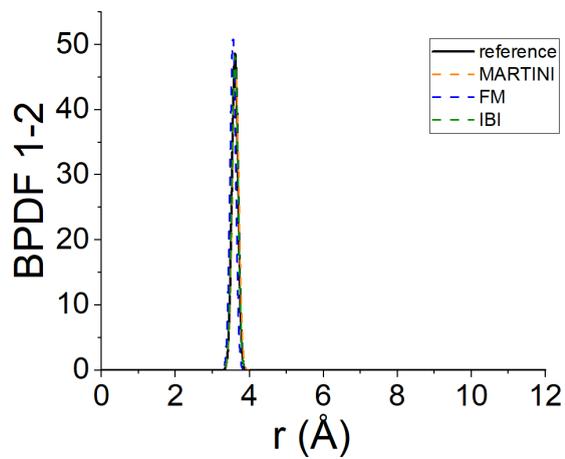
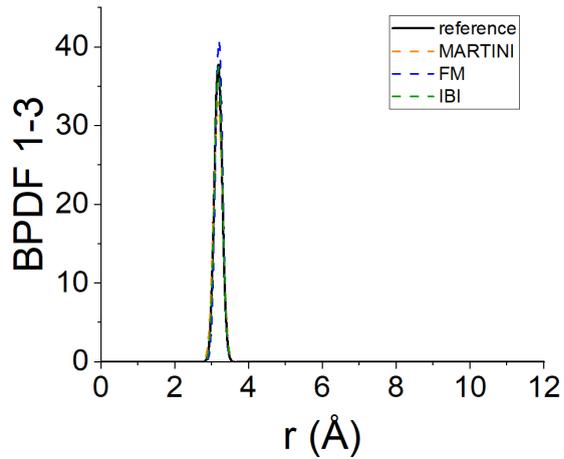
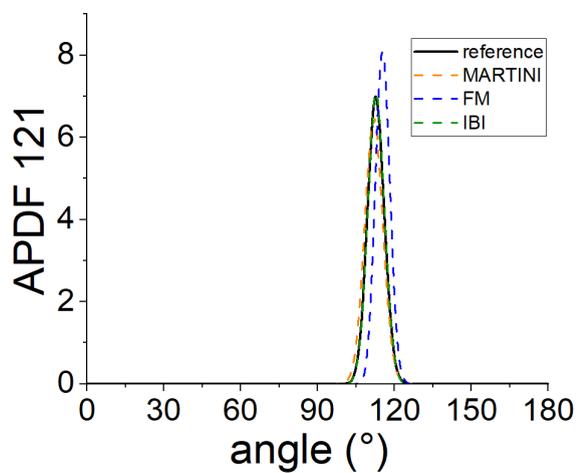
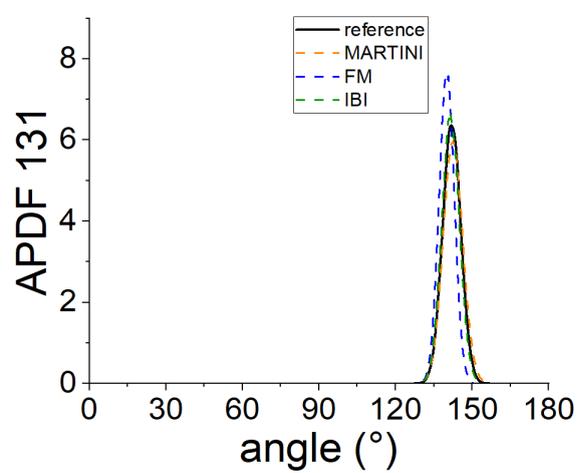
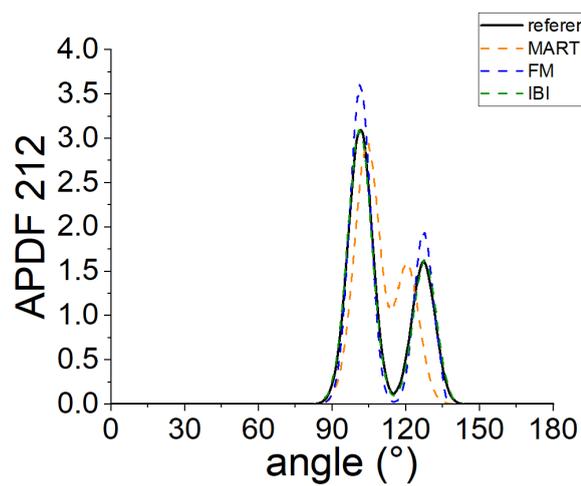
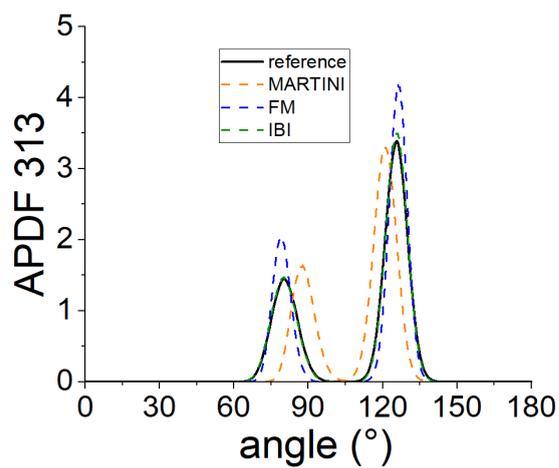

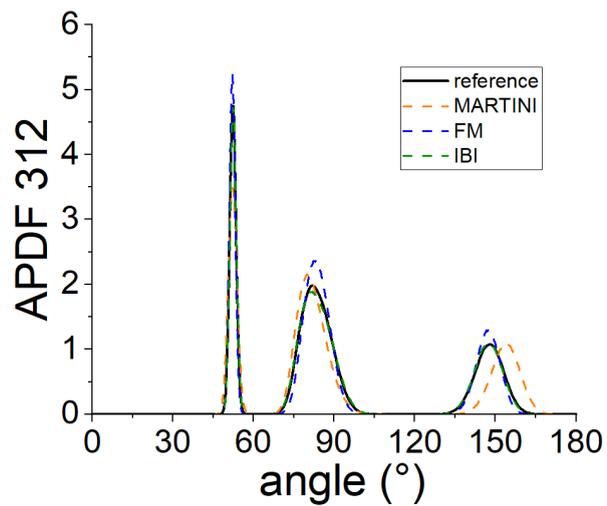
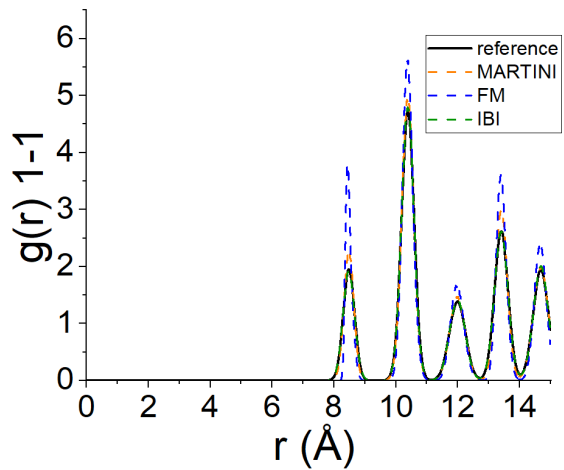
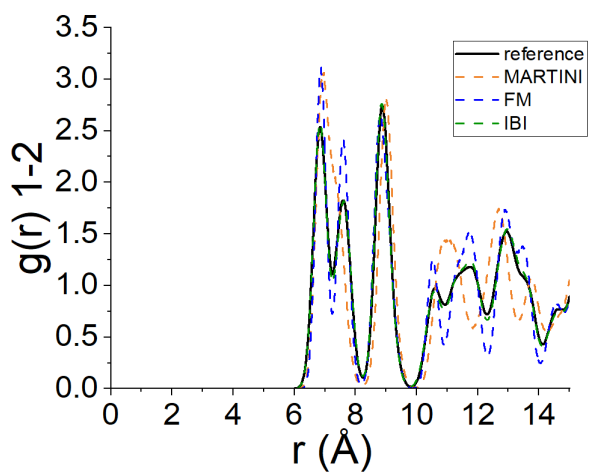
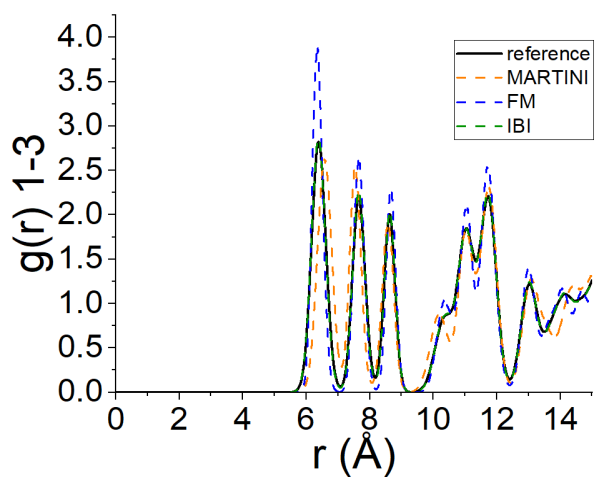
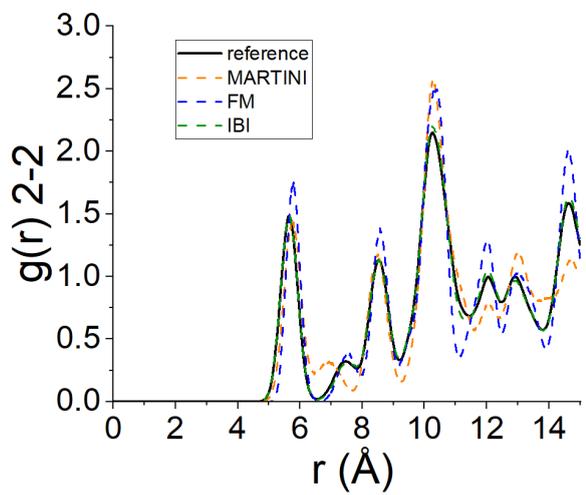
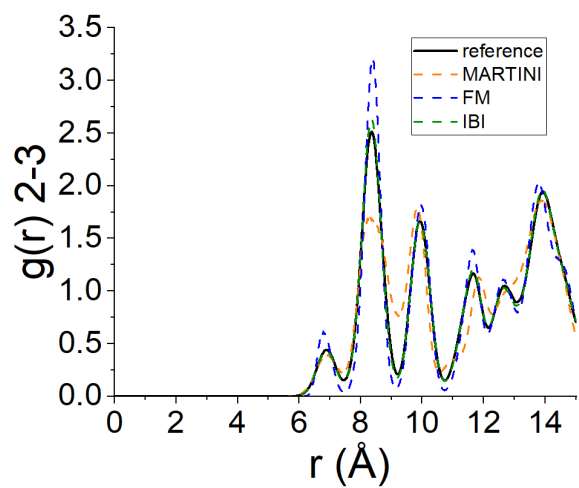

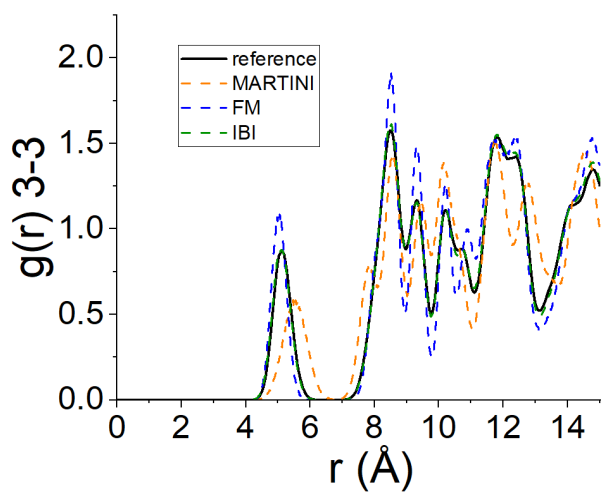

Figure SM5 Structural results given by MARTINI (dashed orange line), FM (dashed blue line) and IBI (dashed green line) models for mapping C. Bond and angle probability density functions are referred to as BPDF and APDF respectively. Reference distributions (solid black line) are also shown.